\documentclass[aps,twocolumn,superscriptaddress,showpacs,floatfix]{revtex4-1}
\usepackage{amsmath,amssymb,graphicx}

\usepackage[utf8]{inputenc}
\usepackage[T1]{fontenc}
\usepackage{xcolor}

\IfFileExists{newtxtext.sty}
   {\usepackage{newtxtext,newtxmath}}
   {\IfFileExists{stix.sty}
      {\usepackage{stix}}
      {\IfFileExists{mathptmx.sty}
      {\usepackage{mathptmx}}{} } }

\usepackage{textcomp}

\usepackage{bm}

\IfFileExists{siunitx.sty}{\usepackage{booktabs,siunitx}}{}

\pdfoutput=1
\usepackage{color}
\definecolor{LinkColor}{rgb}{0.256,0.439,0.588}
\usepackage{hyperref}
\hypersetup{
   pdfauthor={Xiao Yan Xu, Yang Qi, Junwei Liu, Liang Fu and Zi Yang Meng},
   pdftitle={Self Learning Determinantal Quantum Monte Carlo Method},
   colorlinks=true,
   citecolor=LinkColor,
   linkcolor=LinkColor,
   urlcolor=LinkColor
}

\usepackage{pifont}

\begin{document}

\title{Self-Learning Determinantal Quantum Monte Carlo Method}

\author{Xiao Yan Xu}
\affiliation{Beijing National Laboratory for Condensed Matter Physics and Institute
of Physics, Chinese Academy of Sciences, Beijing 100190, China}
\author{Yang Qi}
\affiliation{Department of physics, Massachusetts Institute of Technology, Cambridge, MA 02139, USA}
\author{Junwei Liu}
\affiliation{Department of physics, Massachusetts Institute of Technology, Cambridge, MA 02139, USA}
\author{Liang Fu}
\affiliation{Department of physics, Massachusetts Institute of Technology, Cambridge, MA 02139, USA}
\author{Zi Yang Meng}
\affiliation{Beijing National Laboratory for Condensed Matter Physics and Institute
of Physics, Chinese Academy of Sciences, Beijing 100190, China}

\begin{abstract}
Self-learning Monte Carlo method~\cite{liu2016self,liu2016fermion} is a powerful general-purpose numerical method recently introduced to simulate many-body systems. In this work, we implement this method in the framework of determinantal quantum Monte Carlo simulation of interacting fermion systems. Guided by a self-learned bosonic effective action, our method uses a cumulative update~\cite{liu2016fermion} algorithm to sample auxiliary field configurations quickly and efficiently. We demonstrate that self-learning determinantal Monte Carlo method can reduce the auto-correlation time  to as short as one near a critical point, leading to $\mathcal{O}(N)$-fold speedup. This enables to simulate interacting fermion system on a $100\times 100$ lattice for the first time, and obtain critical exponents with high accuracy.
\end{abstract}

\date{\today}
\maketitle

{\it Introduction}\,---\,Determinantal quantum Monte Carlo (DQMC)~\cite{Blankenbecler1981,Hirsch1985,AssaadEvertz2008}, as an unbiased numerical method, has been widely used in the investigation of strongly correlated fermionic systems. Successful applications span through  broad arena including the understanding of charge, magnetic, superconductivity and (quantum) critical properties of Hubbard and $t$-$J$ models~\cite{Hirsch1983,Hirsch1985,Preuss1994,Assaad1996,Brunner2000,Staudt2000,Varney2009,Meng2010,Toldin2015,Otsuka2016}, interaction effects on topological state of matter and novel quantum phase transitions in metals~\cite{Hohenadler2012,Assaad2013,Meng2014,He2016a,Wu2016,He2016b,Assaad2016,Gazit2016,Berg2012,
Schattner2015a,Schattner2015b,Xu2016,Assaad2016,Xu2016b}.

Given all these successes in the past three decades, DQMC -- the work horse for investigation of correlated fermion systems -- still face serious difficulties. In DQMC, one introduces bosonic auxiliary fields to decouple the fermion interactions~\footnote{In some cases~\cite{Berg2012,Schattner2015a,Schattner2015b,Xu2016,Assaad2016,Gazit2016}, the bosonic fields already exist in the original Hamiltonian, and their fluctuations mediate effective fermion interactions}, and the weight of each bosonic field configuration is determined by integrating out the fermions, which gives rise to a determinant and inevitably leads to heavy-duty matrix operations. Thus, even the fastest algorithm available still has the computational complexity at least of $\mathcal{O}(\beta N^{3})$\cite{loh2005numerical}, where $\beta$ is the inverse temperature and $N$ is the system size. Moreover, this algorithm has to employ local updates and the generated configurations are usually strongly correlated. Especially, around phase transition points, the autocorrelation time $\tau$ in local updates is very large and dramatically increases with the linear system size. 
Together, these severe scaling behaviors seriously limit the extrapolation of DQMC results to the thermodynamic limit. 

Recently, we proposed a new general-purpose method, self-learning Monte Carlo (SLMC), to speed up MC simulations \cite{liu2016self,liu2016fermion}. Very encouragingly, with highly efficient cumulative update algorithm, SLMC can generally reduce the computational complexity and dramatically decrease the autocorrelation time in fermion systems\cite{liu2016fermion}. In cumulative updates, configurations are updated much more cheaply according to the simple effective bosonic Hamiltonian self-learned in SLMC, instead of the the original Fermion Hamiltonian, 
and heavy-duty matrix operations are hence avoided.
At the same time, the final results are guaranteed to be statistically exact by the detailed balance principle obeying the original Hamiltonian.

In this work, we extend SLMC to fermionic quantum many-body systems in the framework of DQMC, hereby referred as SLDQMC. Using cumulative update scheme, SLDQMC manage to greatly reduce the computational complexity of the original DQMC to at least a factor 
$\mathcal{O}(N)$ for fixed $\beta$, hence, the larger the system sizes, 
the higher speedup of SLDQMC over DQMC. Moreover, cumulative update is a highly efficient global update algorithm, which effectively reduce the autocorrelation time to be $\mathcal{O}(1)$ around phase transition points, independent of system size. These advantages make it possible now to access larger system sizes. For example, in this work, we are able to simulate interacting fermionic model at $(2+1)$d space-time with system size $20\times100\times100$, where $20$ is the temporal dimension and $100$ is the spatial dimension, a number unaccessible in the previous literatures.

{\it Basics of DQMC}\,---\,To set the stage for SLDQMC, we need to first briefly introduce DQMC. Let's start with the partition function of a general fermionic quantum many-body system
\begin{equation}
Z = \sum_{\{\mathcal{C}\}} \phi(\mathcal{C}) \det \left( \mathbf{1} + \mathbf{B}(\beta,0;\mathcal{C}) \right),
\end{equation}
where $\mathcal{C}=\{s_{i,\tau}\}$ is the auxiliary field ($s_{i,
\tau}$) configuration after the Hubbard–Stratonovich (HS) transformation is applied to decouple the fermion interaction terms in the Hamiltonian~\cite{AssaadEvertz2008} or the bosonic filed already involved in the original model~\cite{Berg2012}. The imaginary time $\beta$ is divided into $M$ time slices ($M\Delta \tau = \beta$) and hence the
configurations $\mathcal{C}$ of the bosonic fields have both spatial and
temporal dependence. 
$\phi(\mathcal{C})$ is
the bare part (including the transformation constant) of the bosonic field, and it is a scalar function. Now for each auxiliary field configuration, the fermions are non-interacting and can hence be traced out, resulting in a determinant $\det \left( \mathbf{1} + \mathbf{B}(\beta,0;\mathcal{C}) \right)$. The matrix $\mathbf{B}(\beta,0)$, depending on configurations $\mathcal{C}$, is a short form for the matrix product  $\mathbf{B}^M \mathbf{B}^{M-1} \cdots \mathbf{B}^1$,  where the matrix at time slice $\tau$ is $\mathbf{B}^\tau = \exp({\Delta \tau \mathbf{K}})\exp({\mathbf{V}(s_{i,\tau})})$, with $\mathbf{K}$ the tight-binding hopping matrix of the bare system in the single-particle basis, and $\mathbf{V}(s_{i,\tau})$ the fermion interaction part after HS transformation , it describes the coupling between bosonic field and fermion bilinear~\cite{AssaadEvertz2008}. The dimension of matrix $\mathbf{B}^\tau$ is equal to the number of degrees of freedom of fermion and usually scale with system size $N \sim L^d$, with $L$ the linear system size and $d$ the spatial dimension.

To update the auxiliary field configuration in DQMC, one performs local update~\cite{Blankenbecler1981,Hirsch1985,AssaadEvertz2008}, i.e., try to flip the bosonic spins $s_{i,\tau}$ one by one through the space-time lattice $\beta N$. The acceptance ratio of such update involves a ratio of two determinants before and after the flip. The computational complexity for evaluating a determinant is $\mathcal{O} \left( N^3 \right)$, but the local nature of the update enables one to perform a fast update with complexity $\mathcal{O} \left( 1 \right)$ to calculate the ratio and complexity $\mathcal{O} \left( N^2 \right)$ to update Green's function if the local update is accepted. However, since one needs to scan over the space-time lattice -- this is called one sweep -- to attempt flip the $\beta N$ numbers of auxiliary field, thus the local update of DQMC is of the computational complexity $\mathcal{O} \left( \beta N^3 \right)$.

There is another factor that further increase the computational complexity -- general for all the Monte Carlo simulation -- the autocorrelation time $\tau_L$. In the context of DQMC, $\tau_L$ is the number of sweeps one needs to perform to have two statistically independent configurations, such that Monte Carlo measurements can be taken.
Therefore, the total computational complexity in DQMC is $\mathcal{O} \left( \beta N^3 \tau_L \right)$. 
At (quantum) critical points or when there are strong correlations in the auxiliary field, the autocorrelation time usually becomes very large and will scale with system size $\tau_L \sim L^z$, which is referred as critical slowing down and $z$ is the dynamic exponent of MC simulation. For local update, $z$ could be very large ($\ge 2$).
In the classical and quantum spin or bosonic systems, tailor-made global update schemes, such as the Swendsen-Wang~\cite{SwendsenWang1987}, Wolff~\cite{Wolff1989}, loop and directed-loop~\cite{Evertz2003,Syljuaasen2002}, have been designed, and the dynamic exponent $z$ can be greatly reduced. But these global update schemes are very model-dependent, and in the framework of DQMC, there is still no practical global update available. 

{\it Formalism of SLDQMC}\,---\,To overcome these problems, 
we design SLDQMC as a general-purpose solution to fermionic quantum Monte Carlo simulations. 
Below we describe its procedure in four steps. 

At step (i), we use the local update of DQMC to generate enough configurations according to the original Hamiltonian. At step (ii), we try to obtain an effective model by self-learning~\cite{liu2016self,liu2016fermion}. The effective model can be very general,
\begin{equation}
H^{\text{eff}} =  E_0 +  \sum_{ (i\tau);(j,\tau')}J_{i,\tau;j\tau'} s_{i,\tau} s_{j,\tau'}+\cdots
\label{eq:effectiveHam}
\end{equation}
where $J_{i,\tau;j\tau'}$-s parameterize the two-body interaction between any bosonic field in space-time. More-body interactions, denoted as $\cdots$, can also be included. In practice, we can use symmetries (rotation, translation, etc) to reduce the number of independent interactions.  We introduce a parameter $\gamma$ as the range of the interactions considered in the effective model, which we can tune to achieve a balance between the accuracy and efficiency.

\begin{figure}[htp!]
\includegraphics[width=\columnwidth]{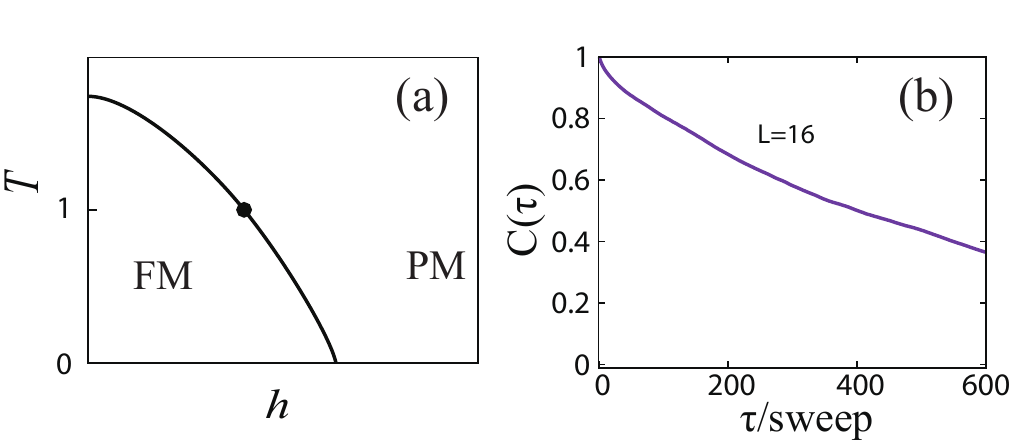}
\caption{(a) Schematic phase diagram of the transverse field Ising model couples to Fermi surface. As a function of the transverse field, the system (both fermions and Ising spins) goes through a transition from ferromagnetic (FM) metal to paramagnetic (PM) metal. The black dot is the finite temperature critical point $(T=1, h_c=2.774(1))$ where we systematically demonstrate the superior performance of SLDQMC over DQMC. (b) Autocorrelation function $C(\tau)$ for $L=16$ system at the critical point in (a), for local update with DQMC, the autocorrelation time is very long (larger than $600$ sweeps).
The autocorrelation function is defined as $C(\tau) = \left( \langle M(0)M(\tau) \rangle -\langle M\rangle ^2 \right) /
\left( \langle M^2 \rangle - \langle M \rangle^2 \right) $ with $M(\tau)$ the total magnetization of Ising spins for the $\tau$-th sweep. }
\label{fig:phase_and_auto}
\end{figure}

The training procedure is straightforward. Given a configuration $\mathcal{C}$ and corresponding weight $\omega[\mathcal{C}]$, generated in the step (i), we have
\begin{equation}
-\beta H^{\text{eff}} [\mathcal{C}] = \ln \left(  \omega[\mathcal{C}] \right).
\label{eq:fitting}
\end{equation}
Combine Eq.~\ref{eq:effectiveHam} and Eq.~\ref{eq:fitting}, optimized values of $\{J_{i,\tau;j\tau'}\}$ can be readily obtained through a multi-linear regression ~\cite{liu2016self,liu2016fermion} using all the configurations prepared in the step (i). 

The effective Hamiltonian can be viewed as an approximation to the exact Hamiltonian for the bosonic field $\mathcal{C}$ after integrating out the fermions, $H[\mathcal{C}]=-\frac{1}{\beta}\ln \omega[\mathcal{C}]$, it contains the low-energy fluctuations of the fermionic degree of freedom in the orginal Hamiltonian. By construction, the Boltzmann distribution of $H^{\text{eff}}[\mathcal{C}]$ approximately reproduces the desired distribution $\omega[\mathcal{C}]$ for dominant configurations of $H[\mathcal{C}]$.

At step (iii) of SLDQMC, we perform multiple local updates with $H^{\text{eff}}$ (as in general the $H^{\text{eff}}$ will contain non-local term which makes the cluster update difficult to implement). Different from the local update in DQMC~\cite{Blankenbecler1981,Hirsch1985,AssaadEvertz2008}, the local move of $H^{\text{eff}}$ is very fast, as there are no matrix operations involved. Furthermore, to generate statistically independent configurations at (quantum) critical point, we need to perform about $\tau_L$ sweeps of local update. With these local updates of effective model, the configuration has been changed substantially, and we take the final configuration as a proposal for a global update for the original model and this entire process is denoted as a {\it cumulative update}. The acceptance ratio of the cumulative update can be derived from the detail balance as
\begin{equation}
 A(\mathcal{C} \rightarrow \mathcal{C}') = \min \left\{1, \  \frac {\exp\left(-\beta H[\mathcal{C}'] \right) } { \exp\left(-\beta H[\mathcal{C}]\right)}  \frac {\exp\left(-\beta H^{\text{eff}} [\mathcal{C}] \right)} {\exp\left(-\beta H^{\text{eff}} [\mathcal{C}'] \right)}  \right\},
\label{eq:acceptanceratio}
\end{equation}
here one clearly see, the closer the $H^{\text{eff}}$ to the original Hamiltonian $H$ with fermion integrated out, the larger $A(\mathcal{C} \rightarrow \mathcal{C}')$ becomes, and eventually, for a good enough $H^{\text{eff}}$, $A(\mathcal{C} \rightarrow \mathcal{C}') \sim 1$ can be achieved in all practical terms (will show below). At step (iv),  following this detailed balance decision, we decide to accept or reject the final configuration. By repeating step (iii) and (iv), we can simulate the interaction fermion systems with high accuracy and efficiency.

\begin{figure}[htp!]
\includegraphics[width=\columnwidth]{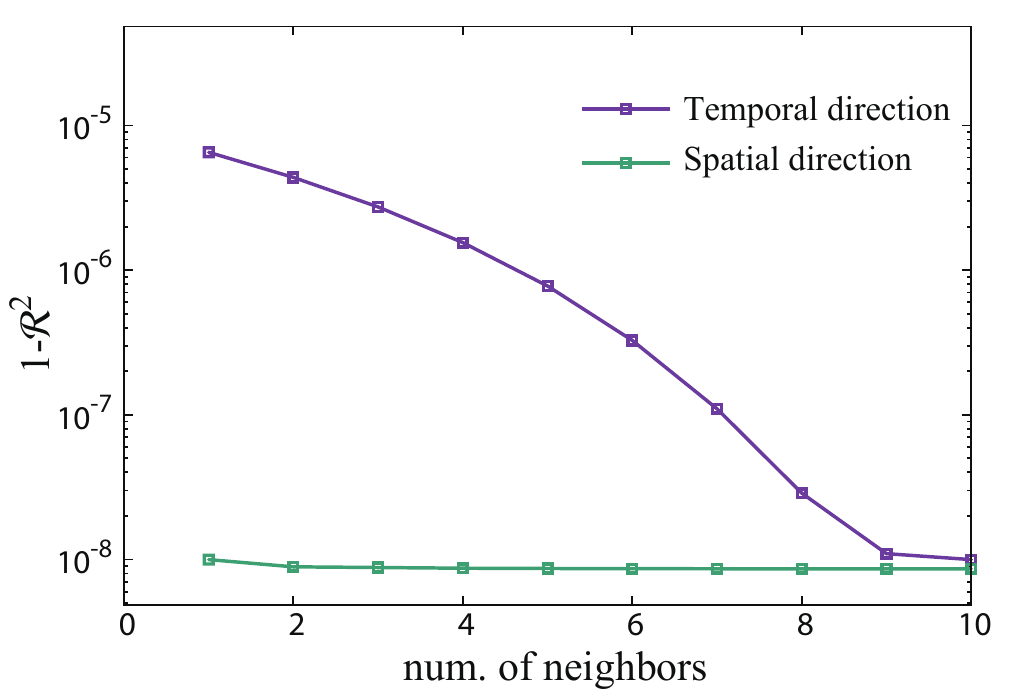}
\caption{ The coefficient of determination for multilinear regression $\mathcal{R}^{2}$ decides how many $J_{i,\tau;j,\tau'}$s (how many neighbor interactions) need to be considered in the $H^{\text{eff}}$. Purple line shows the $1-\mathcal{R}^{2}$ for the temporal neighbors while fix the spatial neighbor the nearest. Green line shows the $1-\mathcal{R}^{2}$ for the spatial neighbors while fix the temporal neighbors. The temporal interaction is more long-ranged (up to the 10th nearest neighbor) but the spatial interaction is short-ranged (up to the 2nd nearest neighbor).}
\label{fig:ediff_and_auto}
\end{figure}

Before we reveal the results of SLDMQC, let's discuss the enormous speedup of SLDQMC over DQMC. The complexity of the cumulative update in SLDQMC is $\mathcal{O} \left( \gamma \beta N \tau_L + \beta N^2 + N^3 \right)$ and it is comprised of two parts. First, the operation to update the effective model is $\mathcal{O} \left( \gamma \beta N \tau_L \right)$. $\gamma$ is the number of operations needed for a single local update on effective model and there are $\beta N$ bosonic field in total, and one performs $\tau_L$ sweeps on all the space-time bosonic fields. Second, the complexity of calculating the acceptance ratio in Eq.~\ref{eq:acceptanceratio} is $\mathcal{O} \left( \beta N^2 + N^3 \right) $. $\beta N^2$ comes from the evaluation of matrix $\mathbf{B}(\beta,0;\mathcal{C})$, 
in that,
$\mathbf{B} (\beta,0;\mathcal{C})$ is the product of $\mathcal{O}(\beta)$ number of $\mathbf{B}^\tau$ matrices, each $\mathbf{B}^\tau$ is a product of $\mathcal{O}(N)$ number of sparse matrices while
the complexity of the dense and sparse matrices production here is $\mathcal{O}(N)$. The other $\mathcal{O}\left( N^3 \right) $ comes from the complexity of calculating the determinant $\text{det}\left( \mathbf{1} + \mathbf{B}(\beta,0;\mathcal{C}) \right)$.

Comparing the $\mathcal{O} \left( \gamma \beta N \tau_L + \beta N^2 + N^3 \right)$  of SLDQMC and $\mathcal{O} \left(\beta N^3 \tau_L \right)$ of DQMC, we define a speedup factor $\mathcal{S}$ of SLDQMC over DQMC and find
\begin{equation}
\mathcal{S} = \text{min}\left(\frac{N^2}{\gamma}, N\tau_L,\beta\tau_L\right).
\label{eq:speedup}
\end{equation}
For many models, we only need include short-range interactions in the effective model \cite{liu2016self,liu2016fermion}, and in this case, SLDQMC can easily reduce the computational complexity by at least of $\mathcal{O} \left( N \tau_L \right)$ or $\mathcal{O} \left( \beta \tau_L \right)$, i.e. the larger the systems and the lower temperature, the more speed up SLDQMC gains. Moreover, it is clear that SLDQMC with cumulative update effectively renders the autocorrelation time to only one sweep, and hence fully cures the critical slowing down at (quantum) critical points.
At last, it is worth noting that even in the worst case, where we need to take long-range interactions in $H^{\text{eff}}$ into account, $\gamma \sim \beta N $, a large speedup, $\mathcal{S} = \mathcal{O} (N/\beta)$ can still be guaranteed, i.e. for given temperature $\beta$, we can achieve at least $\mathcal{O} (N)$-fold speedup.
All those advantages make SLDQMC very suitable to study the interacting fermion systems with large sizes, especially around critical points.




\begin{figure}[tp!]
\includegraphics[width=\columnwidth]{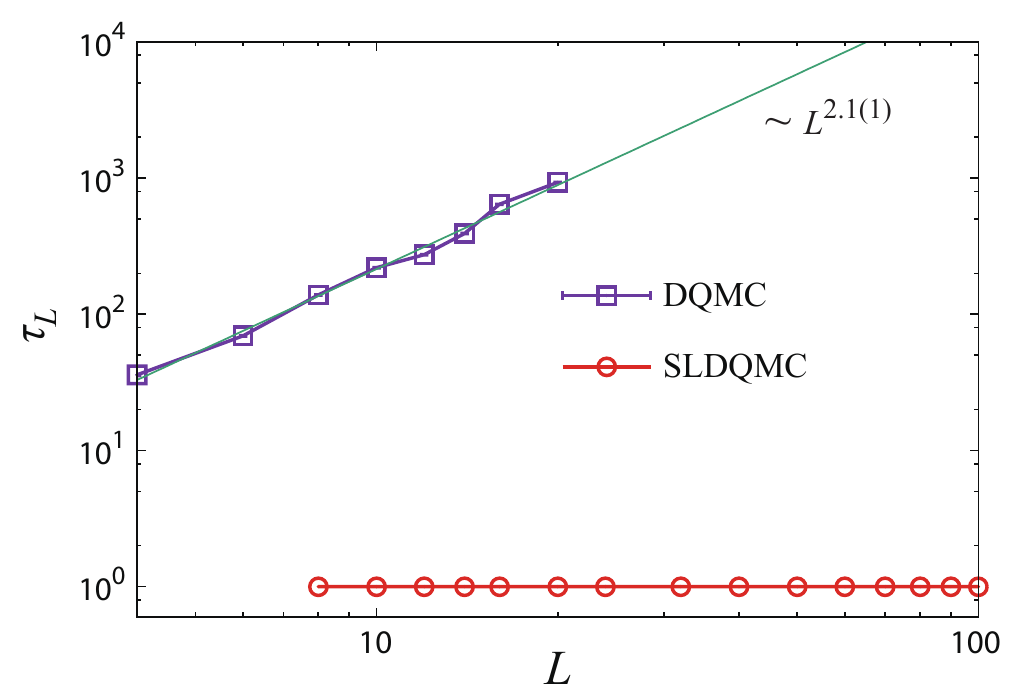}
\caption{Comparison of $\tau_L$ between DQMC and SLDQMC at the critical point. For DQMC, the critical slowing down with $\tau_{L}\sim L^{2.1(1)}$ is observed, while for SLDQMC, the critical slowing down has been complete cured, $\tau_{L}=1$ for all the system size up to $L=100$. 
}
\label{fig:auto_scaling}
\end{figure}

{\it Results}\,---\,To demonstrate the power of SLDQMC, we consider an interacting fermion model with ferromagnetic transverse-field Ising spins coupled to a Fermi surface. The Hamiltonian is comprised of three parts,
\begin{equation}
H=H_f +  H_{s} + H_{sf}.
\label{eq:hamiltonian}
\end{equation}
The fermion part, $H_{f}=-t\sum_{\langle ij\rangle\lambda\sigma}(c_{i\lambda\sigma}^{\dagger}c_{j\lambda\sigma}+h.c.)-\mu\sum_{i\lambda\sigma}n_{i\lambda\sigma}$, describes spin-1/2 fermion hoppings on a bilayer ($\lambda=1,2$) square lattice, with intralayer hopping $t$ and chemical potential $\mu$. The Ising spin part, $H_{s}=-J\sum_{\langle ij\rangle}s_{i}^{z}s_{j}^{z}-h\sum_{i}s_{i}^{x}$, with ferromagnetic $J$ and transverse field $h$ introducing (quantum) fluctuations to the system. At $T=0$, the Ising spins go though a quantum phase transition from ferromagnetic (FM) phase to paramagnetic (PM) phase at $h_c/J=3.04$ with $(2+1)$d Ising universality~\cite{Pfeuty1971,BatrouniScalettar2011,Xu2016}, and at finite temperature, the transition from FM to PM is of 2d Ising universality. The $H_{sf}=-\xi\sum_{i}s_{i}^{z}(\sigma_{i1}^{z}-\sigma_{i2}^{z})$ is the coupling between Ising spin and fermion spin, the coupling favors a parallel (antiparallel) alignment of Ising spin and fermion spin in layer 1 (2). Such bilayer setup guarantees a sign-problem-free QMC simulation in the framework of DQMC~\cite{Xu2016}.

Once switching on the coupling, $\xi=1$, the fluctuations in the Ising spins introduce effective interaction to the fermions, and the fermions will in term introduce long-range interactions among the Ising spins. Our model in Eq.~\ref{eq:hamiltonian} thus provides an ideal situation to study the behavior of itinerant electrons with quantum fluctuations in the vicinity of (quantum) critical points in a controlled manner. Such itinerant quantum critical point (FM-QCP in this case) is at the heart of strongly correlated electron systems. 
In particular, the question of Fermi-liquid instabilities at magnetic quantum phase transition~\cite{Stewart2001,Loehneysen2007} and its applications to heavy-fermion materials and transition-metal alloys such as cuprates and pnictides, is of vital importance and broad interest to condensed matter physics community.

The quantum phase transition of our model in Eq.~\ref{eq:hamiltonian} and the properties of the fermions in the quantum critical region is still not fully understood to date. Theoretical approaches able to address such problems are still understand intensive development~\cite{Chubukov2004,Chubukov2009,Senthil2008,Lee2009,Dalidovich2013,Fitzpatrick2014}.
And numerically evidences are being collected and supporting the universality here is different from the usual $(2+1)$d Ising class~\cite{Xu2016b}.

\begin{figure}[tp!]
\includegraphics[width=\columnwidth]{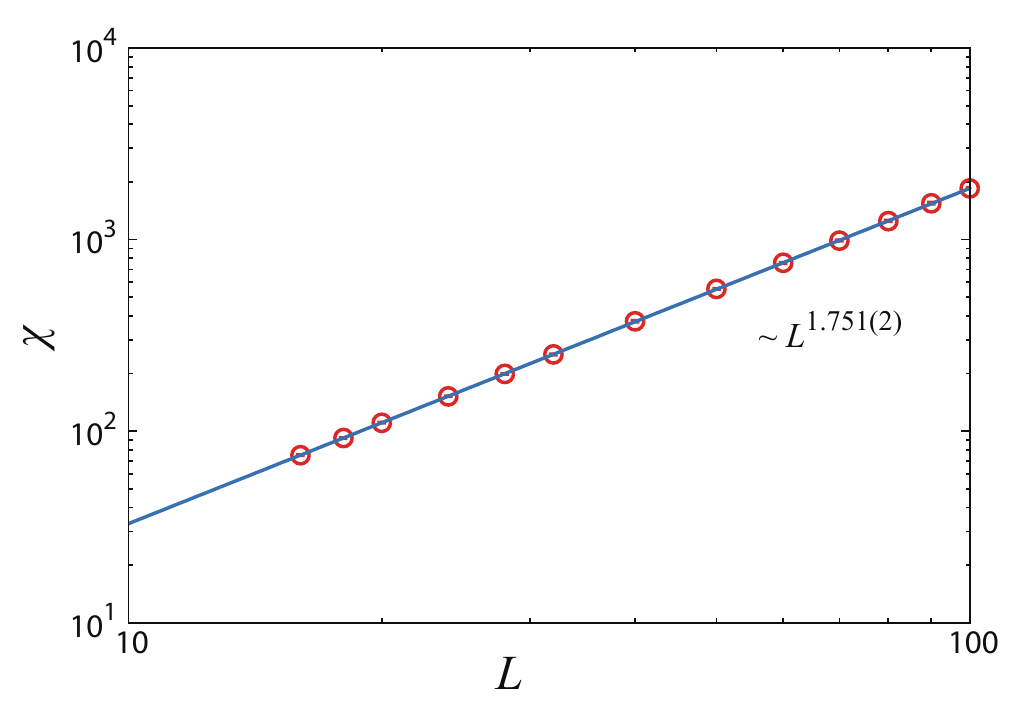}
\caption{Uniform spin susceptibilities $\chi$ at the critical point as a function of system sizes, $\chi \propto L^{2-\eta}$ with $\eta=\frac{1}{4}$ as the anomalous dimension of 2d Ising universality. The linear system size is as large as $L=100$.}
\label{fig:largesize}
\end{figure}

Although the quantum critical properties are complicated, finite temperature FM to PM phase transition is relative simple and one can have a simple schematic phase diagram as shown in Fig.~\ref{fig:phase_and_auto} (a). In this work, we demonstrate the power of SLDQMC by focusing on FM-PM critical point at a finite temperature: $\beta=1.0$, $\Delta\tau=0.05$, $M=20$ and $h_c=2.774(1)$, as the black dot in Fig.~\ref{fig:phase_and_auto} (a). At the critical point, configurations of Ising spins generated by local updates become strongly correlated, as shown by the autocorrelation function of Ising spin in Fig.~\ref{fig:phase_and_auto} (b) for $L=16$, it is the exact manifestation of critical slowing down.

To train the effective model in Eq.~\ref{eq:effectiveHam}, we use $1-\mathcal{R}^2=\langle ( H^{\text{eff}}- H )^{2}\rangle/(\langle H^2\rangle-\langle H \rangle^2)$, where $\mathcal R^2$ is the coefficient of determination (as a figure of merit "score") for the multilinear regression in Eq.~\ref{eq:fitting}. Fig.~\ref{fig:ediff_and_auto} shows the $1-\mathcal{R}^2$ of the multilinear regression as we vary the range of interactions in the effective model. For the purple line in Fig.~\ref{fig:ediff_and_auto}, we fix the $J_{i,\tau;j,\tau}$ to only nearest-neighbor in spatial direction and explore the range of the interaction in the temporal direction. It turns out that the interaction in the temporal direction is long-ranged, and since we choose $M=20$ time slices in total, one needs to consider the interaction up to $M=10$. The spatial range of interaction, on the other hand, is short-ranged. As shown with green line  in the Fig.~\ref{fig:ediff_and_auto}, we keep the interaction in the temporal direction to $M=10$ and plot of the $1-\mathcal{R}^2$ as a function of spatial range, one can clearly see that after the 2nd nearest-neighbor, the $1-\mathcal{R}^2$ is already converged to  a very small value. In the real fitting, when let the range of interactions in both temporal and spatial direction free, we find for $L=8$ system at $\beta=1.0$, in total 16 $J_{i,\tau;j,\tau'}$-s (2 spatial neighbors, 10 temporal neighbors, 4 spatial-temporal neighbors) in $H^{\text{eff}}$ are needed to give the best fit at the critical point.



With effective model obtained from $L=8$, we now perform SLDQMC with cumulative update for larger system sizes.
The great improvement is shown in Fig.~\ref{fig:auto_scaling}. The autocorrelation time of DQMC presents the typical critical slowing down behavior: $\tau_{L} \propto L^{z}$ with $z=2.1(1)$. However,  SLDQMC overcomes such a slowing down completely: $\tau_{L}$ is basically a constant as small as one for all the system sizes simulated, and the dynamic exponent of SLDQMC with cumulative update  is practically $z=0$. Due to such superior behavior of SLDQMC, a speedup of $\mathcal{S}=\mathcal{O}(N)$ for our 2d system is easily achieved, as promised in the discussion of Eq.~\ref{eq:speedup}. 

With such a speedup by SLDQMC, we are now able to access enormously large system. In Fig.~\ref{fig:largesize}, we measure the uniform Ising spin susceptibility
$\chi (L) = \frac {1} {\beta L^2} \sum_{ij} \int_0^\beta d\tau \langle s_{i,\tau}^z s_{j,0}^z \rangle$.  Since the system is at a 2d Ising critical point, $\chi(L) \propto L^{2-\eta}$, with $d=2$ and $\eta=\frac{1}{4}$. We are able to simulate systems as large as $L=100$ and $L^{2-\eta}$ is clearly seen in the $\chi$ with $2-\eta=1.751(2)$. We would like to emphasize that, this is for the first time, a $(2+1)$d interacting fermionic system with space-time dimension $20\times100\times100$, has ever been simulated in quantum Monte Carlo simulations, which is only possible by using SLDQMC.

{\it Discussion}\,---\,In this Letter, we extend SLMC method~\cite{liu2016self,liu2016fermion} to the fermionic quantum many-body systems, and implemented it in the framework of DQMC. The hence obtained SLDQMC, with cumulative update scheme, provides a general purpose solution to fermionic quantum Monte Carlo simulations. We demonstrate that SLDQMC can reduce the autocorrelation time to as short as one sweep at the critical point, and speed up the simulation at least of $\mathcal{O}(N)$-fold at fixed temperature. To illustrate the strength of SLDQMC, a 2d interacting fermion system with size as $100\times100$ is for the first time, being able to simulated. We believe, SLDQMC with cumulative update, opens a new avenue for the numerical investigation of interacting fermionic system. After three decades of intensive studies with DQMC, it is now possible to simulate system size as large as those in the QMC study of quantum spin systems. Many standing problems in the interacting fermion system are now able to reveal in certainty with SLDQMC.

\section*{Acknowledgments}
X.Y.X and Z.Y.M. acknowledge the support from the Ministry of Science and Technology (MOST) of China under Grant No. 2016YFA0300502, the National Natural Science Foundation of China (NSFC Grants No. 11421092 and No. 11574359), and the National Thousand-YoungTalents Program of China. The work at MIT is supported by the DOE Office of Basic Energy Sciences, Division of Materials Sciences and Engineering under Award DE-SC0010526. LF is partly supported by the David and Lucile Packard Foundation. We thank the following institutions for allocation of CPU time: the Center for Quantum Simulation Sciences in the Institute of Physics, Chinese Academy of Sciences; the National Supercomputer Center in Tianjin and the Gauss Centre for Supercomputing e.V. (www.gauss-centre.eu) for providing access to the GCS Supercomputer SuperMUC at Leibniz Supercomputing Centre (LRZ, www.lrz.de).
\appendix

\bibliography{main}

\end{document}